\documentclass{aastex63}
\usepackage{gensymb}
\begin{document}
\title{A stellar constraint on Eddington-inspired Born-Infeld gravity from cataclysmic variable binaries} 
\correspondingauthor{Tapobrata Sarkar}
\email{tapo@iitk.ac.in}

\author{Pritam Banerjee}
\affiliation{Department of Physics, Indian Institute of Technology,
	Kanpur 208016, India}

\author{Debojyoti Garain}
\affiliation{Department of Physics, Indian Institute of Technology,
	Kanpur 208016, India}

\author{Suvankar Paul}
\affiliation{ICFAI University, Agartala, Tripura 799210, India}

\author{Rajibul Shaikh}
\affiliation{Department of Physics, Indian Institute of Technology,
	Kanpur 208016, India}

\author{Tapobrata Sarkar}
\affiliation{Department of Physics, Indian Institute of Technology,
	Kanpur 208016, India}

\begin{abstract}
Eddington-inspired Born-Infeld gravity is an important modification of Einstein's general relativity, which can 
give rise to non-singular cosmologies at the classical level, and avoid the end-stage singularity in a gravitational
collapse process. In the Newtonian limit, this theory gives rise to a 
modified Poisson's equation, as a consequence of which stellar observables acquire model dependent corrections,
compared to the ones computed in the low energy limit of general relativity. 
This can in turn be used to establish astrophysical constraints on the theory. 
Here, we obtain such a constraint using observational data from 
cataclysmic variable binaries. In particular, we consider the tidal disruption limit of the secondary star by a white dwarf primary. 
The Roche lobe filling condition of this secondary star is used to compute stellar observables in the modified gravity 
theory in a numerical scheme. These are then contrasted with the values obtained by using available data on these objects, 
via a Monte Carlo error progression method. This way, we are able to constrain the theory within $5\sigma$ confidence level. 
\end{abstract}

\section{Introduction}
\label{sec1}

In spite of the unprecedented success of Einstein's general relativity (GR), modifications thereof are important, and
have been the focus of intense research over the last few decades. 
On one hand, issues related to the observed cosmic acceleration and the cosmological constant indicate that
such modifications are possibly necessary. From a more fundamental point of view however, theories beyond GR might be
essential, due to the biggest {\it pathology} of GR itself, namely the existence of mathematical singularities, 
such as the big bang. In GR, the process of matter collapsing
under its own gravity often leads to unavoidable singularities, indicating the limitation of the theory itself. 
Although it is commonly believed that quantum effects might smoothen
these, a consistent theory of quantum gravity has been elusive. An alternative then is to construct
a singularity-free classical theory of gravity itself. One can envisage such a theory in lines with the celebrated works
of \cite{Born1}, \cite{Born2}, \cite{BI1}, who constructed a viable theory of electromagnetism free of the divergences associated 
with the more conventional Maxwell form. 

One such theory that has been the focus of attention in the recent past is the Eddington inspired Born-Infeld (EiBI) theory
of gravity, put forward by \cite{BF} (BF), building upon the work of \cite{DG}, \cite{Vollick1, Vollick2}. 
Briefly put, BF used an alternative to the Einstein-Hilbert
action of GR, proposed by Eddington, where the gauge connection is considered as a fundamental field as opposed
to the metric tensor (\cite{Eddington}, \cite{Schrodinger}). In Eddington's formalism, the gravitational Lagrangian,
apart from pre-factors is taken to be $\sqrt{|{\rm det} R_{\mu\nu}|}$, with $R_{\mu\nu}$ being the Ricci tensor. 
Variation of this action with the affine connection being considered as a dynamical variable 
(dubbed as the Palatini formalism) leads to Einstein's 
equation in the presence of a cosmological constant, which is otherwise obtained from a variation of the more conventional
Einstein-Hilbert Lagrangian proportional to $\sqrt{|{\rm det}g_{\mu\nu}|}(R - 2\Lambda)$.
Here, $R = g^{\mu\nu}R_{\mu\nu}$ is the Ricci scalar,
$\Lambda$ is the cosmological constant, and the metric $g_{\mu\nu}$ is the dynamical variable. 
BF considered a Born-Infeld type of ``square root'' action, with a minimal coupling of gravity with matter
fields, with a Lagrangian proportional to $\sqrt{|{\rm det}[g_{\mu\nu} + \epsilon R_{(\mu\nu)}]|}-\lambda
\sqrt{|{\rm det}g_{\mu\nu}|}$, apart from the matter contribution, and one considers the 
symmetric part of the Ricci tensor in the Lagrangian, denoted by the braces. Here, $1/\epsilon$ is the 
Born-Infeld mass $M_{BI} \ll M_{Pl}$, the Planck mass, and $\lambda$ is a dimensionless non-zero parameter,
related to the cosmological constant, with asymptotically flat solutions corresponding to $\lambda = 1$. 
BF showed that this theory led to singularity-free cosmology. 
Shortly afterwards, \cite{Pani1} showed that EiBI gravity is indeed capable of avoiding singularities that arise due
to a collapse process, see also \cite{Delsate1}. 

In the non-relativistic limit, EiBI gravity gives rise to a modified Poisson's equation, with the modification 
of the low energy limit of Einstein gravity being
characterised by a coupling term that is non-zero only in the presence of matter. Since the Poisson's equation
is used as a basic input in many of stellar observables, it is then natural that EiBI theories can thus be tested
by stellar physics. Indeed, there has been a variety of works in the recent past in this direction. \cite{Casanellas1}
proposed tests for the theory using solar constraints. \cite{Avelino1} studied such constraints using cosmological
and astrophysical scenarios. \cite{Avelino2} obtained bounds on EiBI theories by demanding that electromagnetic
forces dominate gravitational ones in nuclear reaction. More recently, \cite{TP} have put constraints on the theory
from an analysis of white dwarfs. \cite{Olmo1} studied gravitational waves in non-singular EiBI cosmological models. 
A recent comprehensive review of EiBI gravity and related phenomenological tests appear in the work of \cite{OlmoRev}. 

The starting point of phenomenological studies of EiBI gravity in astrophysical scenarios is the modified
Poisson's equation that the theory yields in the Newtonian limit. With the speed of light denoted by $c$
and the gravitational constant by $G$, expanding the field equations up to first order in $\epsilon$,
one obtains (\cite{BF}, \cite{OlmoRev}),
 \begin{equation}
 \nabla^2 \phi = \frac{8 \pi G}{2c^4} \rho + \frac{8\pi G \epsilon}{4c^4}\nabla^2 \rho~,
 \label{eq.Poisson1}
 \end{equation}   
where $\epsilon = 1/M_{BI}$. In units $c=G=1$, we will write this equation as 
\begin{equation}
 \nabla^2 \phi = 4\pi \rho +\frac{\kappa_g}{4}\nabla^2 \rho~,
 \label{eq.Poisson}
 \end{equation}
with $\kappa_g = 8\pi\epsilon$. 
Observables in stellar physics, in which the Poisson's equation is a crucial input, therefore gets modified in EiBI gravity,
and observational data can be used to constrain $\kappa_g$. In this paper, we will use data
from tidal forces in cataclysmic variable (CV) binary systems (for a comprehensive introduction to CV systems,
see \cite{CVbook}) to put such a constraint. As we will discuss in details 
in the next section, the fact that the donor star in such a binary fills up its Roche lobe provides us with
a way to compute all the observables (like critical mass, radius, etc.) and Eq. (\ref{eq.Poisson}) then 
implies that these are dependent on the EiBI parameter. Then, using catalogued data on these 
binaries allows us to put bounds on $\kappa_g$. We are able to provide 
$5\sigma$ bounds on this parameter. 

Importantly, an attractive feature of Eq. (\ref{eq.Poisson}) is that the modification to the Poisson's equation does not
require us to assume spherical symmetry. This is in contrast with many other known therories where
such modifications take place. For example, in the beyond-Horndeski class of modified gravity theories
(for a recent review, see \cite{KobayashiRev}),
only the radial part of the modified pressure balance equation is known, and it is imperative in such examples 
to assume spherical symmetry of stellar objects. This assumption has to be carefully dealt with, in 
CV systems where the stellar structure is more complicated due to effects of tidal forces as well
as rotations, as discussed by \cite{tapo3}. EiBI theories on the other hand provide a much neater 
picture where one can work explicitly in Cartesian coordinates. 

In this paper, we will be interested in studying EiBI gravity in the context of low mass main sequence stars that are well
described by a polytropic equation of state that relates the pressure to the density, and is of the form 
$P = \kappa\rho^{1+\frac{1}{n}}$, with $\kappa$ being the polytropic constant and $n$ the polytropic index. 
These are the secondaries in CV binary systems with a white dwarf 
primary, and we will focus on stars of mass $\sim 0.4M_{\odot}$. For such low mass stars, the polytropic index
$n = 1.5$ is a good approximation (see, e.g., \cite{polytrope_1}, \cite{polytrope_2}), a fact that is also borne out 
by the observation these stars have a rotation time period of less than $6$ hours. Here we use the 
observational data on CV binaries, with the underlying theory being EiBI gravity. This latter fact is 
inbuilt in our analysis which uses the modified Poisson's equation, Eq.(\ref{eq.Poisson}), and via this,
stellar observables are numerically obtained as a functions of $\kappa_g$, thus offering ready 
comparison with data, which allows us to constrain the possible values of $\kappa_g$. In the next
section \ref{Num}, we will briefly elaborate on the numerical recipe that we use. Section \ref{Meth}, we
discuss the methodology of constraining the EiBI parameter $\kappa_g$ followed by section \ref{sec-3.1}
which contains our main results. The paper ends with a summary of the results in section \ref{Summary}. 

\section{Numerical Procedure}
\label{Num}

We will now discuss the procedure to constrain the EiBI parameter using 
a set of observed data points and comparing it with that obtained numerically. Our numerical procedure involves 
solving the modified Poisson equation and the Euler equation for hydrostatic equilibrium inside a polytropic star. 
We find the star's deformed shape under the influence of tidal forces, such that the star is maximally deformed and 
fills its Roche lobe. Under a stronger tidal field, the star will be tidally disintegrated. Thus, the Roche lobe
filling condition gives the critical mass that the star must have, so that it is not tidally disrupted. 
This is a condition that we can use as a source of information to find the masses and radii of the secondary 
stars in a CV system. Comparing them with the observed data allows us to constrain the EiBI parameter. This
will be the approach that we will follow in this paper.   
Thus we can consider a polytropic fluid star in flat spacetime throughout its complete trajectory. 
The effect of the background curvature now comes in the form of an additional force field, namely the tidal field.
To begin with, we numerically create a polytropic star which remains in hydrostatic equilibrium under the influence of 
a tidal field. The tidal field is calculated in a locally flat Fermi-Normal (FN) frame (\cite{Manasse-Misner}). The FN 
frame is particularly useful, as its local flatness allows us to deal with fluid equations in the Newtonian limit,
and the local inhomogeneity of gravity is already incorporated in the tidal field. 

In the FN frame, the tidal potential can be written as \cite{Ishii-Kerr}: 
\begin{equation}
\phi_{\text{tidal}} =\frac{1}{2} C_{ij} x^i x^j + \frac{1}{6} C_{ijk} x^i x^j x^k + \frac{1}{24} \left[ C_{ijkl} + 
4 C_{\left(ij\right.} C_{\left.kl\right)} - 4 B_{\left(kl|n|\right.} B_{\left.ij\right)n} \right] x^i x^j x^k x^l + O(x^5)
\label{eq.phi_tidal}    
\end{equation}
Here, the FN coordinates are denoted by $ x^i = \{x^0,x^1,x^2,x^3\} $ and the coefficients are given in terms of
the rank $4$ Riemann curvature tensor as
given by 
\begin{equation}
C_{ij} = R_{0i0j}, ~~~ C_{ijk} = R_{0\left(i|0|j;k\right)}, ~~~ C_{ijkl} = R_{0\left(i|0|j;kl\right)}, 
~~~ B_{ijk} = R_{k\left(ij\right)0}~,
\label{eq.Cij}
\end{equation}
where, covariant derivatives are indicated by the symbol ` ; ' and  $ i,j,k,... $ run from $1$ to $3$. 
The tensorial notation $  R_{0\left(i|m|j;kl\right)}$ denotes a summation over all the possible permutations 
of $i,j,k,l$ with $m$ fixed at its position, divided by the total number of such permutations.
In the presence of the tidal force field, the deformation is obtained by numerically solving the Euler equation in the 
FN frame, 
\begin{equation}
\rho \frac{\partial v_i}{\partial \tau} + \rho v^j \frac{\partial v_i}{\partial x^j} = - \frac{\partial P}{\partial x^i} - 
\rho \frac{\partial (\phi + \phi_{\text{tidal}})}{\partial x^i} +\rho \left[ v^j \left( \frac{\partial A_j}{\partial x^i} - 
\frac{\partial A_i}{\partial x^j} \right) - \frac{\partial A_i}{\partial \tau} \right]~.
\label{eq.hydrodynamic}
\end{equation} 
Here, as discussed before, $P=\kappa\rho^{1+(1/n)}$ is the pressure inside the fluid body of the star. The last term on the right hand side 
comes from gravitomagnetic forces where $A_i = \frac{2}{3} B_{ijk} x^i x^j$ is the corresponding vector
potential. $v^i$ is the velocity field of a fluid element, and $\phi_{\text{tidal}}$ is the tidal potential as 
experienced by the star inside the FN frame. The self-gravity of the star is calculated from the modified 
Poisson equation of Eq. (\ref{eq.Poisson}). 

For non-zero $\kappa_g$, the second term in Eq.(\ref{eq.Poisson}) quantifies the difference in the self-gravity of the
star, as compared to GR. 
It is important to note that $\kappa_g$ is expressed in SI units as $m^5 kg^{-1} s^{-2} = [[G]][[ R^2]] $, where 
$G$ is the gravitational constant and $R$ is a length that is usually related to the size of the star. 
This implies that $\kappa_g$ will have a higher value as one considers a star of a larger radius, to remain 
significant in Eq.(\ref{eq.Poisson}) and hence any bound on $\kappa_g$ will be dependent on the particular
star that one considers. A better procedure is then to define the dimensionless quantity 
$\bar{\kappa}_g = \kappa_g / (G R^2)$ which does not depend on the size of the star and therefore 
can constrained using any astrophysical object. Here, the only input in our analysis is the polytropic
equation of state, and hence the bound that we obtain on the EiBI parameter is universal for all low mass main 
sequence stars. 

We solve Eq.(\ref{eq.Poisson}) and Eq.(\ref{eq.hydrodynamic}) to find the central density, mass and the volume equivalent 
radius of the deformed star. The star is corotating in the FN frame with velocity (\cite{Ishii-Kerr},\cite{tapo3})
\begin{equation}
v^i= \Omega \left[-\{x^3 - x_c \sin(\Omega\tau)\},0,\{x^1-x_c \cos(\Omega\tau)\}\right]
\end{equation}
where $\Omega$ is the corotational frequency, and $x_c$ is a constant that arises as the rotational axis deviates 
from the $x^2$ axis due to the star's deformation. It is convenient to convert these equations into dimensionless 
form since we do not know the amount of deformation beforehand. Firstly, we rewrite the equations in coordinate 
$\tilde{x}^i$ which is defined as the coordinate of the star such that the star has no rotation in the tilde frame and 
the tilde frame rotates in the FN frame with angular frequency $\Omega$ along the $x^2$ axis. Such a choice 
allows us to remove the $\tau$ dependent parts from the above equations. Moreover, we need to write 
$\phi_{\text{tidal}}$ and $\phi_{\text{mag}}$ in terms of the tilde coordinates. Next, we convert the coordinates 
into dimensionless form $\tilde{x}^i = p q^i$, where $p$ is a constant to be found iteratively as it converges to 
a prescribed precision and $q^i$ is the dimensionless coordinate. Now, Eqs.(\ref{eq.Poisson}) and 
(\ref{eq.hydrodynamic}) can be written using the dimensionless coordinates as,   
 \begin{equation}
\frac{\Omega^2}{2} p^2\left[(q^1-q_g)^2  + (q^3)^2\right] = \kappa (n+1) \rho ^{1+1/n} + 
\phi + \phi_{\text{tidal}} + \phi_{\text{mag}} + C
\label{eq.dimless_hydrodynamic}
\end{equation}

\begin{equation}  
\frac{1}{p^2}\nabla_q^2 \phi = 4 \pi \rho + \frac{\kappa_g}{4p^2}	\nabla_q^2 \rho
\label{eq.dimless_poisson}
\end{equation}
where, $\nabla_q^2 \equiv p^2 \nabla^2$ is the dimensionless Laplacian, and in Eq.(\ref{eq.dimless_hydrodynamic}) 
we have used the polytropic equation. As discussed earlier,  in units of $c=G=1$, $\kappa_g$ has the dimension of
length squared whereas $p$ has the dimension of length, i.e., $[[\kappa_g]]=[[p^2]]$. Therefore, the right hand 
side of Eq.(\ref{eq.dimless_poisson}) is well defined in dimensionless form. Now, to solve Eqs.(\ref{eq.dimless_hydrodynamic}) 
and (\ref{eq.dimless_poisson}), we need to fix the boundary conditions in order to find the constants 
$C, p$ and $q_g$. These are obtained by fixing the surface of the deformed star at $ (q_s, 0, 0) $ where 
the density is zero. Also, we provide the central density $ \rho=\rho_c $ and $ \partial \rho/\partial q^1 = 0 $ 
at the center of the star which is assumed to be the origin of the tilde coordinates. Although the origin may not 
coincide with the center of the star in the deformed shape, the error is negligible. We solve 
Eq.(\ref{eq.dimless_hydrodynamic}) and (\ref{eq.dimless_poisson}) together to find the solution in hydrostatic 
equilibrium. An initial density distribution, along with an initial $p$ is used in Eq.(\ref{eq.dimless_poisson}) to 
obtain $\phi$ which is then used in Eq.(\ref{eq.dimless_hydrodynamic}) to find the updated value of $p$. 
The updated $p$ is used in Eq.(\ref{eq.dimless_poisson}) and the iteration continues until $p$ converges 
to a desired precision. This procedure is performed for different central density $\rho_c$ until the Roche lobe 
filling condition is satisfied. At this critical condition, $ \partial\rho/\partial q^1 $ at the surface of the star at 
$ (q_s, 0, 0) $ becomes smoothly zero (for more details on the numerical procedure, see \cite{BANERJEE201929}).
 
To perform the numerical procedure, we still need to specify the polytropic constant $\kappa$, polytropic index $n$, 
and the EiBI gravity parameter $\kappa_g$. As was discussed earlier, the polytropic index is set to $n=1.5$ in our case, 
whereas, $\kappa_g$ is varied to put a constraint on it by comparing the numerical results with the observational data. 
On the other hand, $\kappa$ can be found by equating the volume equivalent radius at the Roche limit, obtained 
numerically in GR ($\kappa_g=0$), with the observed radius. The same $\kappa$ is thereafter used to find 
the critical mass and radius due to other non-zero $\kappa_g$ values. Such an assumption is motivated by the fact
that the polytropic constant is determined entirely by stellar hydrodynamics, and thus should remain unchanged 
for different EiBI parameters. For example, in case of white dwarfs (WDs), $\kappa$ is obtained by 
equating electron degeneracy pressure with the carbon atom density. 
In our case, we use the Roche lobe filling condition to find $\kappa$ for which GR is chosen as a reference. 
Such a procedure is essential when no other ways are known to find the polytropic constant beforehand. 
However, many other possible values of $\kappa$ appear because we can use any other radius within the 
observed range to match the numerical result in GR. We will show in the next section how the best possible 
choice of the observed radius is made.

\section{Methodology}
\label{Meth}

As discussed previously, we compare the numerical data with observational data of a set of cataclysmic variable 
systems which are binary systems with a Roche lobe filling secondary star orbiting a WD primary. When a star 
fills its Roche lobe, it is at its critical mass below which it overfills the Roche lobe. The disrupted material is accreted 
by the primary. This Roche lobe filling condition can therefore be used to find the stellar parameters. We utilize this 
Roche fill condition to find the unknown polytropic constant $\kappa$ for which the star's numerical size matches 
with the observed one. 

A set of 13 CV systems is used in this paper. The orbital distances between the primary and the secondary stars in 
these CVs are large enough to safely neglect the rotation of the primary while calculating the tidal field around the secondary. 
It allows us to model the gravitational field of the primary as a Schwarzschild geometry. However, the radius of the 
secondary is about $\sim 0.1$ times the orbital distance, which is significantly large to generate asymmetry in the 
deformed shape of the secondary. Hence, we take up to the fourth order term in the tidal potential. Also, it is safe 
to assume that the secondary moves in a circular orbit since the type of the orbit does not make any significant 
difference in the tidal field when the orbital distance is large. Thus the observed orbital parameters of these CV 
systems can be used to compare with the numerical data for constraining the EiBI gravity parameter. As already 
mentioned, the polytropic index of the secondary star is taken to be  $n=1.5$. The secondary stars in these 
CVs fall in the main sequence category and have small masses $ \sim 0.4 M_\odot $ as can be interpreted 
from their orbital period less than $6$ hours. These stars are known to have highly convective cores and can be 
accurately represented by a polytropic index $n=1.5$.

We find the Mass of the primary ($M_1$), mass ($M_2$) and radius ($R_2$) of the secondary and the orbital 
distance ($a$) using Monte Carlo error progression method from a set of observed input parameters such as orbital 
period ($P$) and inclination angle ($i$), mass ratio ($q$), binary phases at mid ingress and mid egress 
($\Delta\phi_{1/2}$), radial velocities of the primary and the secondary ($K_1$ and $K_2$) and the rotational velocity 
($vsini$) of the secondary star.  A detailed description of the procedure can be found in 
\cite{BT_Mon_1} \cite{V347_Pup_1} \cite{Monte_Carlo_1}. A list of the input parameters are shown in 
Table \ref{table1} and the output parameters 
$M_1$, $M_2$, $a$ and $R_2$, obtained from Monte Carlo error progression, are given in Table \ref{table2}. 
 
We use the observed $M_1$ and $a$ to find the tidal field in FN frame of the secondary. Next, we numerically 
calculate the critical mass ($M_2^{crit}$) and volume equivalent radius ($R_2^{crit}$) of the star at Roche limit 
in the presence of various nonzero values of $\kappa_g$. It is found that both $M_2^{crit}$ and $R_2^{crit}$ 
increase as $\kappa_g$ is increased. Hence, we get a range of critical mass and radius of the secondary 
from the numerical analysis, which is then compared with the observed ranges. In the next section, we analyze 
the results and find a constraint on the EiBI parameter.

\section{Constraining the EiBI gravity parameter}
\label{sec-3.1}

\begin{figure}[h!]
\hspace{0.3cm}
\centering
\centerline{\includegraphics[scale=0.5]{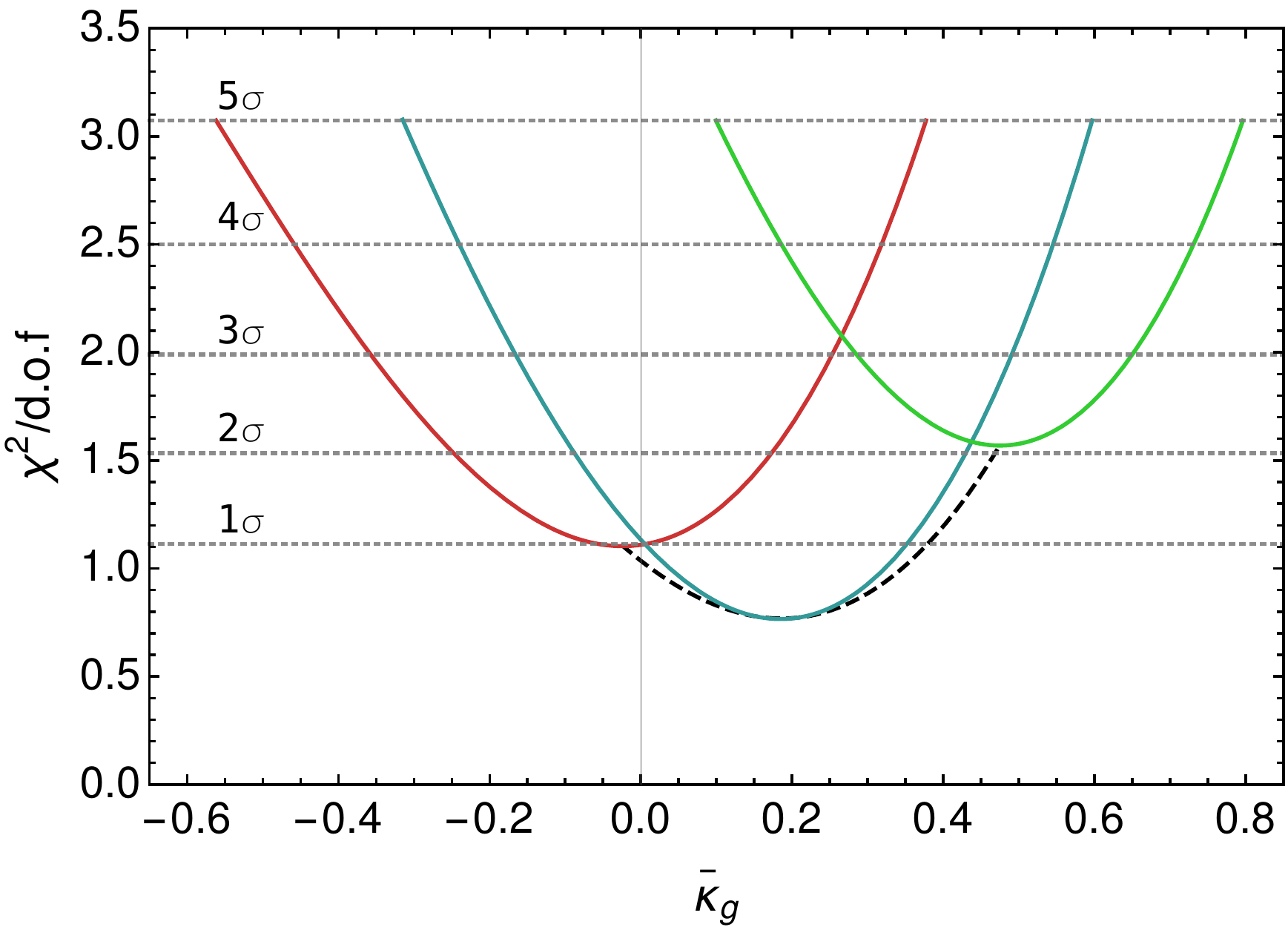}}
\caption{Variation of $\chi^2/\text{d.o.f}$ with $\bar{\kappa}_g$ for different values of $\kappa$ 
which are again obtained for different choices of $R_{2,i}$ such that $R_{2,i}^{crit}=R_{2,i}$ in GR. 
The plot in red is for $R_{2,i}= R_{2,i}^{mean}+\sigma_{R,i}$, in green is for $R_{2,i}= R_{2,i}^{mean}-\sigma_{R,i}$ 
and in blue is for $R_{2,i}= R_{2,i}^{mean}+(\sigma_{R,i}/4)$. The dashed curve in black denotes the 
positions of the minimum points of each individual plots. The blue curve having the lowest minimum, 
is the best choice to constrain $\bar{\kappa}_g$.}
\label{fig:sub1}
\end{figure}
\begin{figure}[h!]
\hspace{0.3cm}
\centering
\centerline{\includegraphics[scale=0.5]{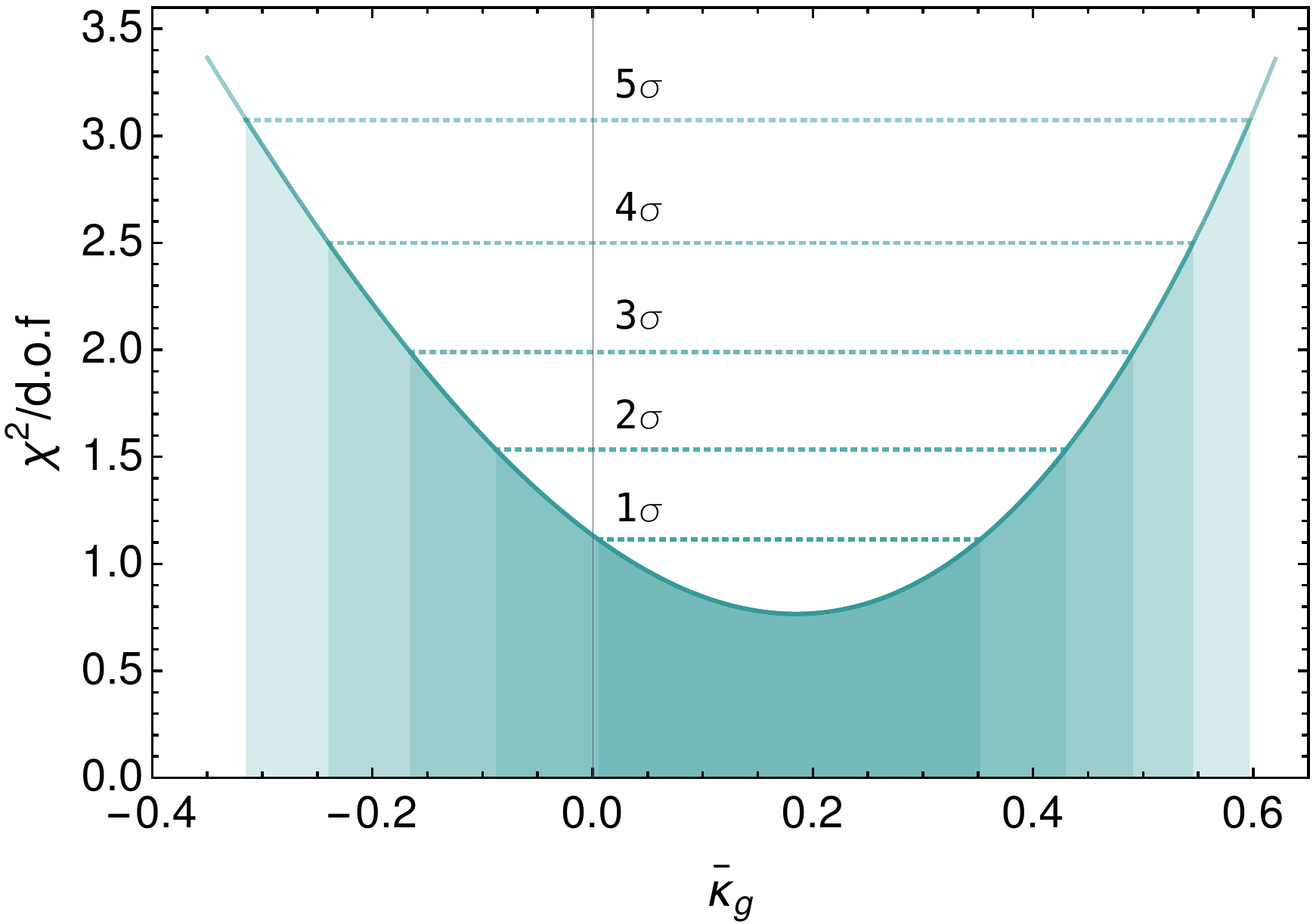}}
\caption{Constrained ranges of $\bar{\kappa}_g$ are shown along with their $\sigma$ levels. }
\label{fig:sub2}
\end{figure}

As the EiBI parameter $\kappa_g$ has a dimension of length squared (in units of $G=c=1$), it is evident that 
the constraint depends on the size of the secondary star $R_2$. However, we can avoid this limitation by constraining 
the dimensionless parameter $\bar{\kappa}_g = \kappa_g/(GR_2^2)$ with any astrophysical object, 
including the CV secondaries. To numerically find the secondary star's critical mass and radius, we need to 
calculate the polytropic constant $\kappa$. The procedure is as follows:

We choose a value of radius $R_2$ of the secondary star from its observed range. Now, we perform the 
numerical procedure keeping $\kappa_g=0$ (i.e., in GR). We can numerically find the volume equivalent 
radius $ R_2^{crit} $ of the star at the Roche limit if any value of the polytropic constant $\kappa$ is given. 
However, the desired value of $\kappa$ is found when $ R_2^{crit} $ becomes equal to $R_2$. 
The same $\kappa$ is then used to numerically calculate the critical masses and radii of the secondary 
for other non-zero values of the EiBI gravity parameter $\kappa_g$. Thus a set of $M_2^{crit}$ and $R_2^{crit}$ 
is obtained numerically for various values of $\bar{\kappa}_g=\kappa_g/(G R_2^{crit})^2$. A similar numerical 
procedure is performed for all the CVs.
 
Now we define a quantity $\chi^2$ as
\begin{equation}
\chi^2 = \sum_{i=1}^{N} \frac{\left(M^{mean}_{2,i}-M_{2,i}^{crit} \right)^2}{\sigma_{M,i}^2} 
+ \frac{\left(R^{mean}_{2,i}-R_{2,i}^{crit} \right)^2}{\sigma_{R,i}^2}~,
 \end{equation} 
where, $M^{mean}_{2,i}$, $\sigma_{M,i}$ and $R^{mean}_{2,i}$, $\sigma_{R,i}$ are the observed mean and 
standard deviation of mass and radius of the secondary star of the $i$th CV system respectively. 
On the other hand, $M_{2,i}^{crit}$ and $R_{2,i}^{crit}$ are the numerically calculated critical mass 
and volume equivalent radius of the Roche lobe filling secondary of the $i$th CV system 
respectively. Here, the total number of systems are taken to be $N=13$. We calculate $ \chi^2 $ 
for various values of $\bar{\kappa}_g$ using the 13 sets of $M_2^{crit}$ and $R_2^{crit}$ 
already obtained numerically. Now, to constrain $\bar{\kappa}_g$ with confidence levels, we need to know 
the degrees of freedom ($\text{d.o.f}$) of the chi-square test. In our case, $ \text{d.o.f}=2N-2=24 $. 
Finally, we find a constraint on $\bar{\kappa}_g$ using the $\chi^2/\text{d.o.f}$ vs. $\bar{\kappa}_g$ plot. 
 
\begin{figure}[h!]
\centering
\includegraphics[scale=0.6]{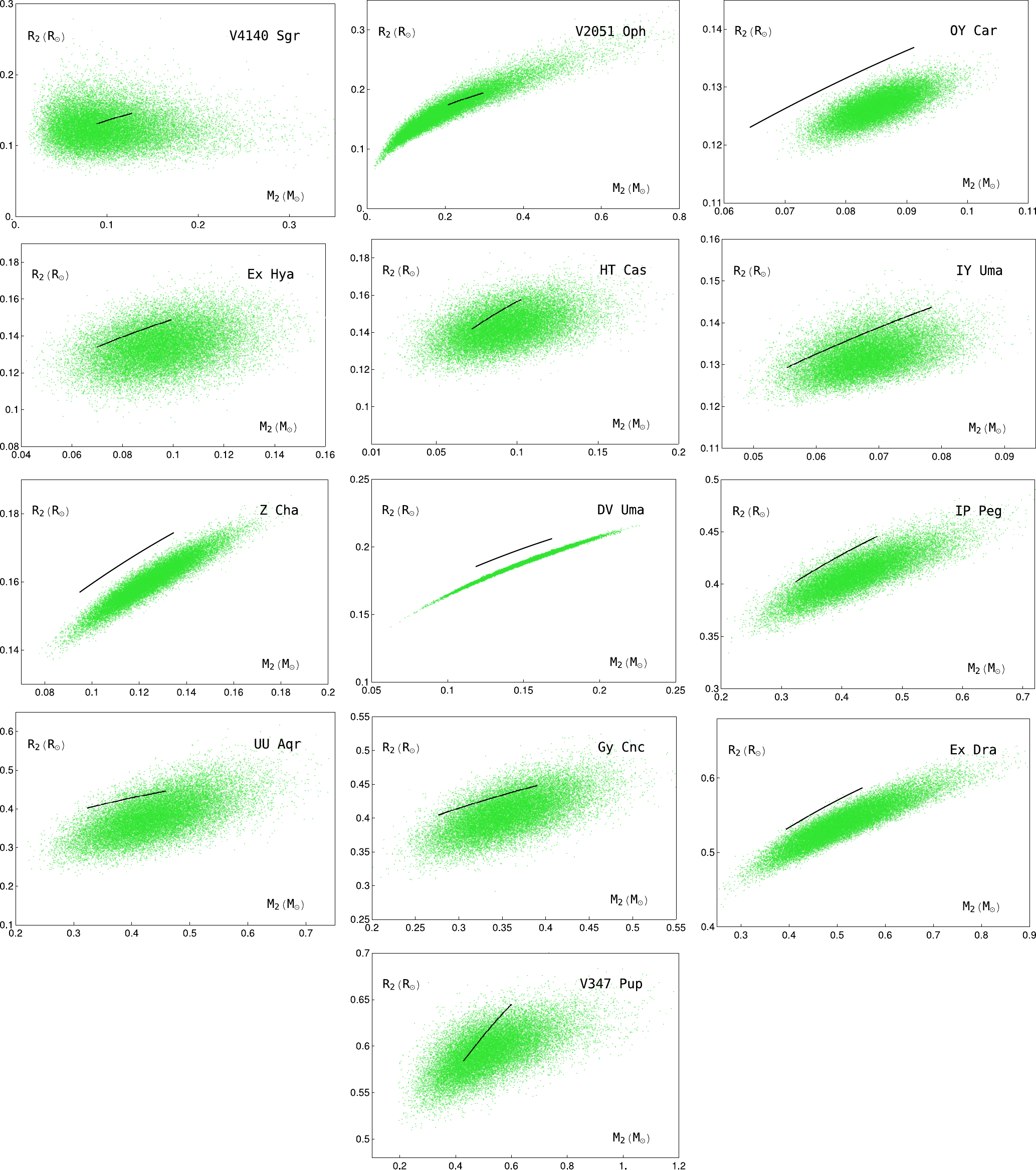}
\caption{In this figure, $M_2$ and $R_2$ ranges of the 13 CV secondaries as obtained from 
Monte Carlo error progression method using the observed parameters given in Table \ref{table1} 
are shown in green dots. These are compared with the numerical values (shown in black lines) 
generated using different values of $\bar{\kappa}_g$ within its $5\sigma$ limits. Higher mass and 
radius appear due to higher values of $\bar{\kappa}_g$.  }
\label{fig2}
\end{figure}

However, this constraint is obtained for choosing a particular set of polytropic constants using a particular 
set of secondary radii from the CV systems. If we take any other set of secondary radii within the 
observed ranges, another set of polytropic constants as well as a different constraint on $\bar{\kappa}_g$ 
is found. The best possible constraint on $\bar{\kappa}_g$ is obtained only if the best set of secondary 
radii is chosen which is done using the chi-square analysis as well. We take different sets of secondary 
stars' radii as $R_{2,i} = R_{2,i}^{mean} + \epsilon \sigma_{R,i}$, where $-1\leq \epsilon \leq 1$. 
Hence, we find different constraints on $\bar{\kappa}_g$ for different choices of $R_{2,i}$. The best choice 
of $R_{2,i}$ is the one for which the minimum of $\chi^2/\text{d.o.f}$ vs $\bar{\kappa}_g$ plot 
has the lowest value. Fig. \ref{fig:sub1} shows three plots between $\chi^2/\text{d.o.f}$ and $\bar{\kappa}_g$ for 
three different sets of $R_{2,i}$. First one is for $R_{2,i} = R_{2,i}^{mean} + \sigma_{R,i}$ (denoted by red) 
meaning the maximum values of $R_{2,i}$, another is for $R_{2,i} = R_{2,i}^{mean} - \sigma_{R,i}$ (green) 
indicating the minimum values of $R_{2,i}$ and the third one stands for $R_{2,i} = R_{2,i}^{mean} 
+ (\sigma_{R,i}/4)$ (blue) for which the $\chi^2/\text{d.o.f}$ vs $\bar{\kappa}_g$ plot has the lowest 
minimum. Therefore, $R_{2,i} = R_{2,i}^{mean} + (\sigma_{R,i}/4)$ is the best choice for calculating 
the desired set of $\kappa$ for each of the CV secondaries and therefore constraining $\bar{\kappa}_g$. 
In Fig. \ref{fig:sub2}, we show the confidence levels to which $\bar{\kappa}_g$ is constrained. We find
 $\bar{\kappa}_g$ to be $0.005 \leq \bar{\kappa}_g \leq 0.352$ within $1\sigma$ and 
 $-0.315 \leq \bar{\kappa}_g \leq 0.597$ within $5\sigma$ confidence level.

In Fig. \ref{fig2}, we show the $M_2$, $R_2$ ranges generated by Monte Carlo using the observed 
parameters as compared to numerically obtained $M_2^{crit}$ and $R_2^{crit}$ which are obtained 
using the best choice of $\kappa$. Numerical data are denoted by black lines which are obtained for 
various values of $\bar{\kappa}_g$ within its $5\sigma$ limit. As already mentioned, both the critical 
mass and radius increase with $\bar{\kappa}_g$. It can be seen that numerical results tend towards 
nonzero positive values of $\bar{\kappa}_g$ to match the observed parameters as also evident from 
the $\chi^2$ analysis. 

There are two important issues that we will discuss at this stage. First, we note that \cite{tapo3}
discussed constraining modified gravity theories of the beyond-Horndeski class. 
An important difference between the method followed there as compared to the present paper
is as follows. In \cite{tapo3}, the polytropic constant is not kept fixed
while varying the modified gravity parameter. As a result, numerical
mass values are influenced by both the polytropic constant as well as the
modified gravity parameter. For each value of this parameter, the
polytropic constant is freely adjusted until the volume equivalent radius
matches the observed radius. While this procedure is correct in its own
merit, it suppresses the modified gravity parameter's effects somewhat, and makes it
difficult to constrain. A better procedure is to consider the modifications
coming solely from the modified gravity parameter. However, eliminating the
effects of the polytropic constant requires its value to be known either
by some other physical equations (like that in the case of WDs) or by
a statistical best choice method. In the case of the CV secondaries, we do not have
any other information to find their polytropic constant beforehand.
Therefore, here we have resorted to the best choice of the polytropic constant
statistically, and we keep it fixed for all values of the modified parameter.

Secondly, we note from Fig. \ref{fig2} that for a few CV systems 
such as OY Car, Z Cha and DV Uma, the numerical data do
not fall onto the Monte Carlo generated distributions. Such tension in the
data appears from the fact that $R_2$ is obtained using
$\Delta\phi_{1/2}$ which is independent of mass $M_2$. On the other hand,
$M_2^{crit}$ is numerically dependent on $R_2$ since it is matched with
the volume equivalent radius of the secondary. Therefore, observed mass
ranges can possibly deviate from the numerical data. We, however, note
that both $M_2^{crit}$ and $R_2^{crit}$ individually fall within the
observed ranges of $M_2$ and $R_2$ respectively while extending both sides
of the observed mean values. That is why, we do not
categorize these CVs as outliers.

\section{Summary}
\label{Summary}

Modifications to GR are becoming increasingly popular of late, as
it is by now commonly believed that such theories might be essential to understand
the nature of gravity at cosmological scales. Apart from this, a significant unresolved issue is
the inevitable singularity at the end of a gravitational collapse process as predicted by
GR. EiBI theories of gravity are very attractive in this sense, as
they give rise to non-singular cosmologies as well as predict non-singular collapse
processes in the realm of classical gravity, i.e., without invoking quantum effects. Although 
the main effects of such modified theories of gravity are expected to set in at strong gravity scales,
nonetheless they often leave their imprint at low energy scales, in this case by a modification
of the Poisson's equation in the Newtonian limit. This allows us to constrain EiBI 
theories via stellar structure tests, and put bounds on the parameter that determines
the deviation from GR. 

To this end, in this paper, we have studied EiBI theories of gravity in the context of CV binaries,
and constrained the theory using available data. A total of $13$ systems were chosen,
with the secondary star orbiting a white dwarf primary and filling its Roche lobe, and being
well described by an $n = 1.5$ polytropic equation of state. 
The Roche lobe filling condition was used to compute stellar observables numerically and 
comparing these with data, we have obtained a constraint on the EiBI parameter $\bar{\kappa}_g$
appearing in Eq.(\ref{eq.Poisson}) which gives, within $5\sigma$ confidence level,
$-0.315 \leq \bar{\kappa}_g \leq 0.597$. 
It is useful to compare this with existing results on stellar bounds of the EiBI parameter 
in the literature, obtained by different methods. \cite{Casanellas1} obtained the bound 
$-0.016 < \kappa_g/(GR_{\odot}^2)<0.013$ from solar physics constraints, and \cite{Avelino1}
obtained an upper bound $\bar{\kappa}_g<4/\pi$ from the fact that the effective
Jeans length in EiBI theories should be less than the solar radius. On the other hand, 
taking a typical white dwarf radius $\sim 10^6$m, the results of \cite{TP} who constrained
EiBI gravity using the mass-radius relation of white dwarfs give, $-0.239 < \bar{\kappa}_g < 0.728$
at $5\sigma$ confidence level.  
 
The only assumption that we have made to simplify the analysis is that of a polytropic 
equation of state inside stellar matter. As we have discussed, this is an excellent 
approximation for low mass CV secondaries considered here, which are fully convective. 
Nonetheless, it might be interesting to relax this assumption and consider a model
with a core-envelope structure. Recently, \cite{tapo4} have explored such a model 
in the context of beyond-Horndeski class of models and studied how modified gravity
affects stellar radius and luminosity in such models. It will be interesting to 
understand these issues in the context of EiBI gravity, and we expect to report on 
this in the near future. 

\newpage
\appendix

In this appendix, we list the observed binary parameters in Table (\ref{table1}) and the stellar
parameters obtained using Monte Carlo error progression in Table (\ref{table2}). These have 
appeared in \cite{tapo3} and are included here for completeness. 

\begin{table}[h!]
\small
\renewcommand{\arraystretch}{1.3} 
\caption{List of observed binary parameters used for calculating $M_1$, $a$, $M_2$ and $R_2$ 
\label{table1}}
\begin{tabular}{cccccccccccccc}
\hline
\hline
Name & \multicolumn{8}{c}{~~Observed binary parameters~~} \\
& $ P $ & $ i $ & $ q $ & $\Delta\phi_{1/2} $
& $ K_1 $ & $ K_2 $ & $ v\sin i $ & $M_2$ &   \\
& (hr) & (degree) &  &  & (km~s$^{-1}$)  &  (km~s$^{-1}$) &  (km~s$^{-1}$) & ($M_\odot$) & \\
\hline
V4140 Sgr$^a$ & $1.467$ & $80.2\pm0.5$ & $0.125\pm0.015$ & $0.0378\pm0.0005$ & $56\pm7$ & - & - & -  \\
V2051 Oph$^b$ & $1.5$ & $83.3\pm1.4$ & $0.19\pm0.03$ & $0.0662\pm0.0002$ & $91\pm12$ & - & -  & - \\
OY Car$^c$ & $1.51$ & $83.3\pm0.2$ & $0.102\pm0.003$ & $0.051\pm0.004$ & - & $470\pm2.7$ & -  & - \\
Ex Hya$^d$ & $1.638$ & $77\pm1$ & - & $0.017\pm0.002$ & $69\pm9$ & $356\pm4$ & - & -  \\
HT Cas$^e$ & $1.77$ & $81\pm1$& - & $0.0493\pm0.0007$ & $58\pm11$ & $389\pm4$  & - & -  \\
IY Uma$^f$ & $1.77$ & $86\pm1$ & $0.125\pm0.008$ & $0.0637\pm0.0001$ & - & $383\pm6$ & - & -  \\
Z Cha$^g$ & $1.79$ & $81.78\pm0.13$ & $0.150\pm0.004$ & $0.0534\pm0.0009$ & - & $430\pm16$  & - & - \\
DV Uma$^h$ & $2.06$ & $84.24\pm0.07$ & $0.151\pm0.001$ & $0.063604$ & - & - & - & $0.15\pm0.02$ & \\
IP Peg$^i$ & $3.797$ & $81.8\pm0.9$ & $0.45\pm0.04$ & $0.0863$ & - & $298\pm8$ & - & -  \\
UU Aqr$^j$ & $3.93$ & $78\pm2$ & - & $0.051\pm0.002$ & $121\pm7$ & $327\pm31$ & - & - \\
Gy Cnc$^k$ & $4.211$ & $77\pm0.9$ & - & $0.060\pm0.005$ & $115\pm7$ & $283\pm17$ & -  &  - \\
Ex Dra$^l$ & $5.04$ & $85^{+3}_{-2}$ & $0.72\pm0.06$ & $0.1085\pm0.0006$ & - & $210\pm14$ & $140\pm10$ & -  \\
V347 Pup$^m$ & $5.566$ & $87\pm3$ & - & $0.115\pm0.005$ & - & $198\pm5$ & $130\pm5$ & -  \\
\hline
\end{tabular}
	 
\footnotesize{$^a$ \cite{V4140_Sgr_1}\cite{V4140_Sgr_2}\cite{V4140_Sgr_3}, $^b$ \cite{V2051_Oph_1}, 
$^c$ \cite{OY_Car_1} \cite{Oy_Car_2} \cite{Oy_Car_3}, $^d$ \cite{Ex_Hya_1} \cite{Ex_Hya_2} \cite{Ex_Hya_3} 
\cite{Ex_Hya_4} \cite{Ex_Hya_5} \cite{Ex_Hya_6} \cite{Ex_Hya_7}, $^e$ \cite{Ht_Cas_1}, $^f$ \cite{IY_Uma_1} 
\cite{IY_Uma_2} \cite{IY_Uma_3}, $^g$ \cite{Z_Cha_1} \cite{Z_Cha_2} \cite{Z_Cha_3}, $^h$ \cite{DV_Uma_1} 
\cite{DV_Uma_2} \cite{DV_Uma_3} \cite{DV_Uma_4}, $^i$ \cite{IP_Peg_1} \cite{IP_Peg_2} \cite{IP_Peg_3} 
\cite{IP_Peg_4}, $^j$ \cite{UU_Aqr_1} \cite{Ex_Hya_5} \cite{UU_Aqr_3}, $^k$ \cite{Gy_Cnc_1}, $^l$ 
\cite{Ex_Dra_1} \cite{Ex_Dra_2} \cite{Ex_Dra_3}, $^m$ \cite{V347_Pup_1} \cite{V347_Pup_2} \cite{V347_Pup_3}
}\\ 
\end{table} 
\newpage
\begin{table}[h!]
\begin{center}
\small
\renewcommand{\arraystretch}{1.3} 
\caption{List of $M_1$, $a$, $M_2$ and $R_2$ as obtained from Monte Carlo error progression method 
using the observed binary parameters given in Table \ref{table1}.  
\label{table2}}
\begin{tabular}{cccccccc}
\hline
\hline
~~~~Name~~~~&~~~~~~~~~~~$M_1 (M_{\odot})$~~~~~~~~~~~&~~~~~$a (R_{\odot})$~~~~~& ~~~~~~~~~~~
$R_2 (R_{\odot})$
~~~~~~~~~~~&~~~~$M_2 (M_{\odot})$~~~~  \\
\hline
V4140 Sgr & $0.9\pm0.5$ & $0.63\pm0.11$ & $0.13\pm0.02$ & $0.10\pm0.05$   \\
V2051 Oph & $1.2\pm0.9$ & $0.726\pm0.14$ & $0.17\pm0.04$ &$0.22\pm0.11$   \\  
OY Car & $1.2\pm0.3$ & $1.48\pm0.11$ & $0.39\pm0.05$ &  $0.085\pm0.003$ \\	
Ex Hya & $0.49\pm0.03$ & $0.589\pm0.014$ & $0.136\pm0.011$  &$0.095\pm0.017$   \\
HT Cas & $0.62\pm0.04$ & $0.661\pm0.018$ & $0.144\pm0.009$  &$0.09\pm0.02$  \\
IY Uma & $0.55\pm0.03$ & $0.630\pm0.011$ & $0.133\pm0.004$ &$0.068\pm0.006$   \\
Z Cha & $0.84\pm0.09$ & $0.7\pm0.3$ & $0.161\pm0.006$  &$0.125\pm0.014$  \\
DV Uma & $1.00\pm0.13$ & $0.86\pm0.04$ & $0.190\pm0.010$  &$0.15\pm0.02$ &  \\
IP Peg & $0.94\pm0.09$ & $1.37\pm0.05$ & $0.41\pm0.02$ &$0.42\pm0.07$ &  \\
UU Aqr & $0.834\pm0.015$ & $0.649\pm0.004$ & $0.127\pm0.0017$ &  $0.44\pm0.07$ \\
Gy Cnc & $0.88\pm0.13$ & $1.42\pm0.07$ & $0.41\pm0.03$ &$0.36\pm0.05$  \\
Ex Dra & $0.69\pm0.10$ & $1.58\pm0.08$ & $0.54\pm0.03$ &$0.52\pm0.09$  \\
V347 Pup & $0.63\pm0.08$ & $1.66\pm0.10$ & $0.60\pm0.02$ &$0.53\pm0.13$   \\							
\hline
\end{tabular}
\end{center}
\end{table}

\end{document}